\documentstyle[twoside,fleqn,epsf,espcrc2]{article}

\epsfverbosetrue

\newcommand{\half}
             {\mbox{\small $\frac{1}{2}$}}          

\newcommand{\AmS}{{\protect\the\textfont2
  A\kern-.1667em\lower.5ex\hbox{M}\kern-.125emS}}

\title{The $\pi N$ Sigma term -- an evaluation using staggered fermions%
           \thanks{The $MT_c$ Collaboration.}%
           \thanks{Talk presented by R. Horsley at LAT93, Dallas, U.S.A.}}
\author{R. Altmeyer%
           \address{Deutsches Elektronen-Synchrotron DESY,
                    Notkestra{\ss}e 85, D-22603 Hamburg, Germany}
        M. G{\"o}ckeler%
           \address{H{\"o}chstleistungsrechenzentrum HLRZ,
                    c/o Forschungszentrum J{\"u}lich, D-52425 J{\"u}lich,
                                                             Germany}
        R. Horsley$^{\rm b,a}$
        E. Laermann%
           \address{Fakult{\"a}t f{\"u}r Physik, Universit{\"a}t Bielefeld,
                    D-33501 Bielefeld, Germany}
        and
        G. Schierholz$^{\rm a,b}$}

\begin{document}

\begin{abstract}
A lattice calculation of the $\pi N$ sigma term is described using dynamical
staggered fermions. Preliminary results give a sea term comparable in
magnitude to the valence term.
\end{abstract}

\maketitle

\setcounter{footnote}{0}

\section{Theoretical Discussion}
\label{theory}

\noindent \raisebox{7.8cm}[0.0cm][0.0cm]{Preprint HLRZ 93-70 \hspace{0.25cm}
                                         November 1993} \hspace{-6.00cm}
The $\pi N$ sigma term, $\sigma_{\pi N}$, is defined as that part
of the mass of the nucleon coming from the expectation value of the up ($u$)
and down ($d$) quark mass terms in the QCD Hamiltonian,
\begin{equation}
   \sigma_{\pi N} = m \langle N | \bar{u}u+\bar{d}d | N \rangle ,
\label{theory.a}
\end{equation}
where we have taken these quarks to have equal current mass ($=m$).
Other contributions to the nucleon mass come from the chromo-electric
and chromo-magnetic gluon pieces and the sea terms due to the $s$ quarks.
Experimentally this matrix element has been measured from low energy
$\pi$-$N$ scattering, \cite{hoehler83a}. A delicate extrapolation
to the chiral limit \cite{cheng88a} gives a result for the
isospin even amplitude of $\Sigma/f_\pi^2$ with $\Sigma = \sigma_{\pi N}$,
from which the $\pi N$ sigma term may be found. The precise value
obtained this way has been under discussion for many years.
For orientation we shall just quote a range of results from later analyses of
$\sigma_{\pi N} \approx 56 \mbox{MeV}$, \cite{gasser88a}, down to
$45 \mbox{MeV}$, \cite{gasser91a}.

To estimate valence and sea contributions to $\sigma_{\pi N}$,
classical current algebra analyses assume octet dominance and make
first order perturbation theory about the $SU_F(3)$ flavour symmetric
Hamiltonian. This gives
\begin{eqnarray}
   \sigma_{\pi N}
              &\approx& m \langle N | \bar{u}u+\bar{d}d -2\bar{s}s
                                    |N \rangle_{sym}    \nonumber \\
              & & \qquad +  2m\langle N |\bar{s}s |N \rangle_{sym}
                                                           \nonumber \\
              &\stackrel{def}{=}& \sigma^{val}_{\pi N} + \sigma^{sea}_{\pi N},
\label{theory.c}
\end{eqnarray}
where we have first assumed that the nucleon wavefunction does not change
much around the symmetric point. We then subtract and add a strange
component. At the symmetric point the $u$ and $d$ quarks each have equal
valence and sea part, while the $s$ quark matrix element only has a sea
component. Thus in the first term the sea contribution cancels, justifying
the definitions given in eq.~(\ref{theory.c}).
Using first order perturbation theory for the baryon mass splittings
$\sigma^{val}_{\pi N}$ may be calculated
to give $\sigma^{val}_{\pi N} \approx 25 \mbox{MeV}$ and so
$\sigma^{sea}_{\pi N} \approx 31 \sim 20 \mbox{MeV}$.
This in turn means that
$ m_s \langle N | \bar{s}s | N \rangle \approx 400 \sim 250 \mbox{MeV}$,
which would indicate a sizeable portion of the nucleon mass
($938\mbox{MeV}$) comes from the strange quark contribution.

\section{Measuring $\sigma_{\pi N}$}
\label{measure}

We now turn to our lattice calculation. We have generated configurations
using dynamical staggered fermions on a $16^3\times 24$ lattice
at $\beta =5.35$, $m=0.01$ (plus some larger masses), \cite{altmeyer93a}.
Practically there are several possibilities open to us for the evaluation
of the matrix element. The easiest is simply to differentiate
the shift operator $\hat{S}_4$ giving
\begin{equation}
   m {{\partial M_N}\over {\partial m}}|_\beta
              = \langle N|m\bar{\chi}\chi|N\rangle = \sigma_{\pi N},
\label{measure.a}
\end{equation}
(the Feynman-Hellmann theorem). Thus we need to measure $M_N$ for
different masses $m$ and numerically estimate the gradient. This leads to
\begin{equation}
   \sigma_{\pi N} \approx 11.9(8)m|_{m=0.01} \approx 0.12(1),
\label{measure.b}
\end{equation}
giving $\sigma_{\pi N} / M_N \approx 0.15$,
which is to be compared with the experimental result of
$\sigma_{\pi N} / M_N \approx 0.06 \sim 0.05$. The numerical result
is much larger, but presumably this simply indicates that we have
used much too large a quark mass in our simulation
($m^{RGI} \approx 35\mbox{MeV}$, \cite{altmeyer93a}).
At present we, and everybody else, cannot avoid doing this.

We now wish to separately estimate the valence and sea contributions.
This is technically more complicated and involves the evaluation of a
$3$-point correlation function, \cite{maiani87a}, ($B=$ baryon)
\begin{equation}
   C(t;\tau) = \langle B(t) m\bar{\chi}\chi(\tau) \bar{B}_W(0) \rangle ,
\label{measure.d}
\end{equation}
($W =$ wall source). This may be diagrammatically sketched as the sum of
two terms:
\begin{figure}[h]
\vspace*{-0.50cm}
\hspace*{2.0cm}
\epsfxsize=3.0cm \epsfbox{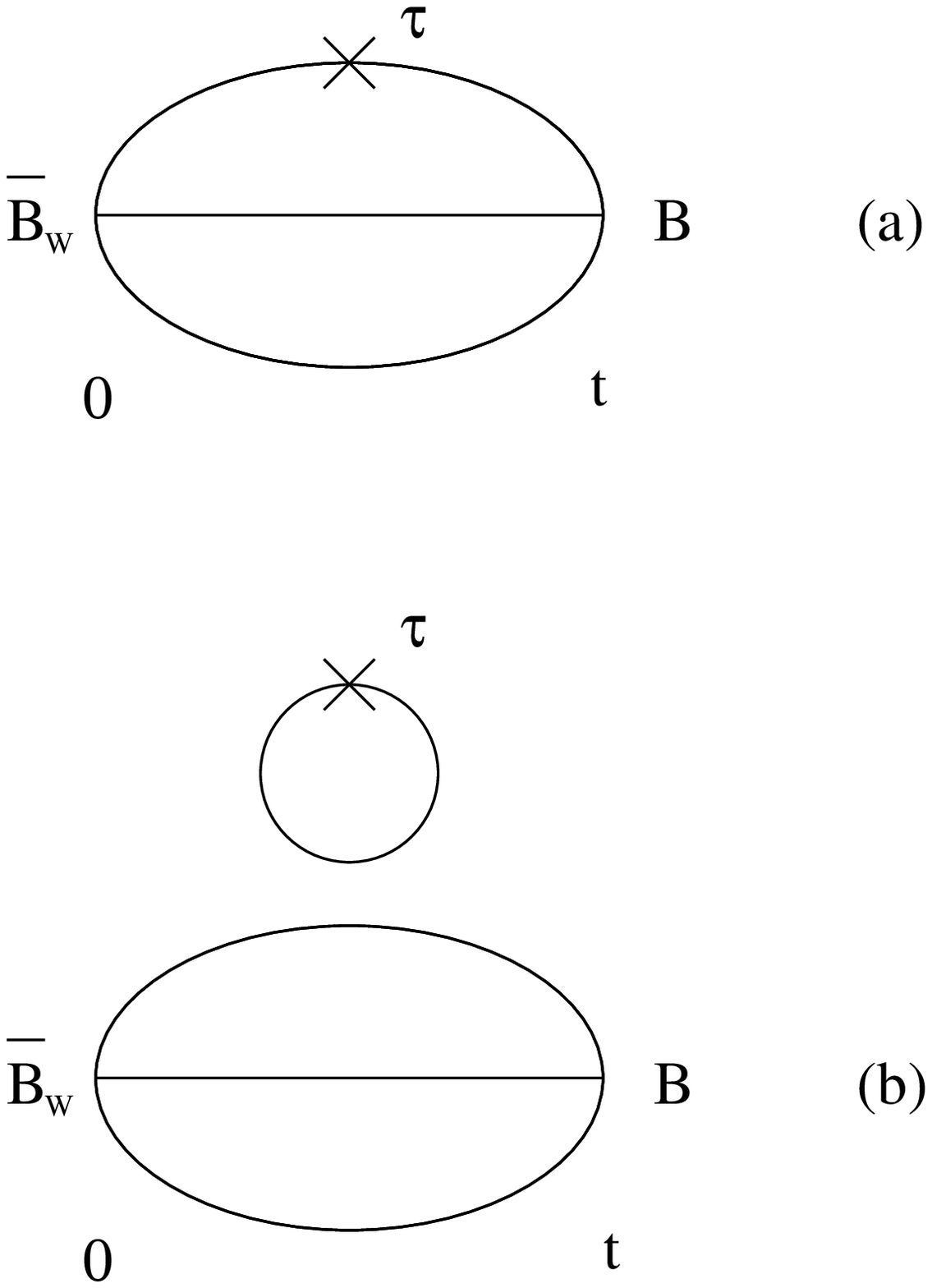}
\vspace*{-0.50cm}
\label{figsketch}
\end{figure}

\noindent
(a) is the connected (or valence) term; the nucleon quark lines are
joined with the mass insertion. (b) represents the disconnected (or sea)
term.

\begin{figure}[thb]
\vspace*{-1.75cm}
\hspace*{-0.50cm}
\epsfxsize=12.5cm \epsfbox{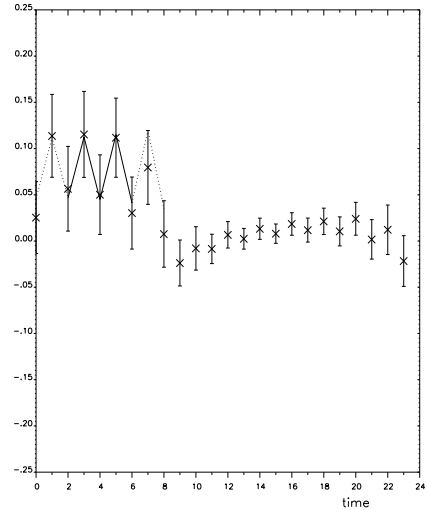}
\vspace*{-1.5cm}
\caption{The ratio of the 3-point connected correlation function to the
         2-point correlation function against $\tau$, with a fit from
         $2$ to $6$.
         (The dotted lines just extend the fit to neighbouring points.)}
\vspace*{-0.50cm}
\label{figconn}
\end{figure}

For the connected piece we have fixed $t$ to be $8$ or $9$
and then evaluated $C(t;\tau)$ as a function of $\tau$. The appropriate
fit function is, with $\mu_\alpha = \xi_\alpha \exp{(-M_\alpha)}$,
($\Lambda$, $\xi_\Lambda=-1$ denoting the parity partner to the nucleon,
$N$, $\xi_N=+1$),
\begin{equation}
   C(t;\tau) \approx \sum_{\alpha, \beta = N,\Lambda} A_{\alpha\beta}
                       \langle \alpha| m\bar{\chi}\chi |\beta\rangle
                       \mu_\alpha^{t-\tau} \mu_\beta^\tau ,
\label{measure.e}
\end{equation}
($\half T \gg t \gg \tau \gg 0$).
$A_{\alpha\beta}$ and $M_\alpha$ are known from $2$-point  correlation
functions. We see that when $\alpha =\beta =N$ we have the matrix element
that we require: $\langle N|m\bar{\chi}\chi|N\rangle$. However
there are other terms which complicate the fit:
$\langle \Lambda|m\bar{\chi}\chi|\Lambda\rangle$ and the cross terms
$\langle \Lambda|m\bar{\chi}\chi|N\rangle \equiv
\mu_\Lambda\mu_N^{-1}\langle N|m\bar{\chi}\chi|\Lambda\rangle$
which are responsible for oscillations in the result. To disentangle the
wanted result from $\langle \Lambda|m\bar{\chi}\chi|\Lambda\rangle$
we need to make a joint fit to the $t=8$ and $t=9$ results. At present
we have not done this, but just checked that separate fits give
consistent results. In Fig.~\ref{figconn} we show preliminary
results for $t=8$. (Up until now we have only evaluated about a quarter
of our available configurations.) We find
\begin{equation}
   \sigma_{\pi N}^{val} \approx 0.08(2),
\label{measure.f}
\end{equation}
with about the same result for the $\Lambda$ matrix element.
The cross term is smaller, roughly $0.02$.

\begin{figure}[htb]
\vspace*{-1.0cm}
\hspace*{-0.50cm}
\epsfxsize=12.5cm \epsfbox{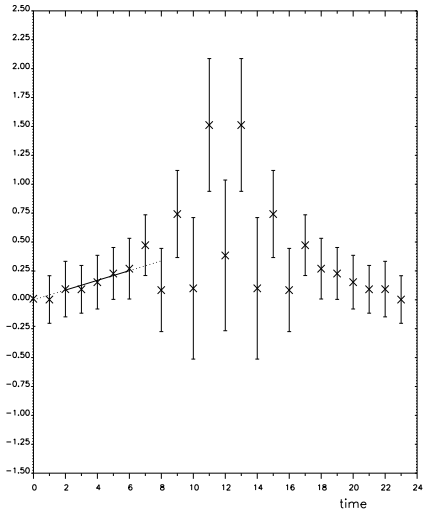}
\vspace*{-1.5cm}
\caption{The ratio for the disconnected correlation function.
         To improve the signal we have averaged
         over the points $t$ and $T-t$.}
\vspace*{-0.75cm}
\label{figdisc}
\end{figure}

Finally we have attempted to estimate the disconnected term
using a stochastic estimator, \cite{bitar89a}.
($100$ sets of Gaussian random numbers were used.) One can improve
the statistics by summing the 3-point correlation function over $\tau$;
this is then equivalent to a differentiation of the 2-point function
with respect to $m$. This gives
\begin{equation}
   \sum_\tau C(t;\tau)
      \approx t \sum_{\alpha=N,\Lambda}
                    A_{\alpha\alpha}
                    \langle\alpha| m\bar{\chi}\chi
                    |\alpha\rangle  \mu_\alpha^t ,
\label{measure.g}
\end{equation}
($\half T \gg t \gg 0$).
In Fig.~\ref{figdisc} we show the disconnected 3-point
correlation function. As expected the quality of the data is poor --
optimistically we see a slope. A simple linear fit gives
\begin{equation}
   \sigma_{\pi N}^{sea} \approx 0.05(4) .
\label{measure.h}
\end{equation}

\section{Discussion}
\label{discussion}

We see that roughly eq.~(\ref{measure.b}) is consistent
with eqs.~(\ref{measure.f},\ref{measure.h}) and that
$\sigma_{\pi N} / \sigma_{\pi N}^{val} \approx 1.5$,
which is to be compared with the experimental value of about
$2.2 \sim 1.8$. Although we can draw no firm conclusions at present
our result tentatively indicates that the valence term
is slightly larger than the sea term.

Comparing our results with other work obtained using dynamical
fermions, \cite{gupta91a} uses $2$ flavours of Wilson fermions and finds
for the ratio $2 \sim 3$, which indicates a somewhat larger sea component.
On the other hand, \cite{patel92a} using results from \cite{brown91a}
for $2$ staggered flavours finds $1.5 \sim 2.0$, while \cite{bernard93a}
has $\sim 2.0$. These, like our result, seem to be lower than for
the Wilson fermion case.

We would also like to emphasise that although lattices can give a
first principle calculation, at present one is not able to do this.
Technically the fit formul{\ae} that we employ, eqs.~(\ref{measure.e},
\ref{measure.g}) are true only for the complete correlation function.
The best way to circumvent this problem and others is
to make simulations at different strange quark masses; differentiation
as in eq.~(\ref{measure.a}) would then give directly
$m_s\langle N|\bar{s}s|N\rangle$, the strange content of the nucleon.
However this calculation is not feasible at the present time.
This should be the ultimate goal of lattice simulations,
as other recent theoretical results,
\cite{gasser91a}, have hinted that perhaps the strange quark content of the
nucleon is not as large as supposed, previous results being explained
by a combination of factors, such as $\Sigma \ne \sigma_{\pi N}$
and higher order corrections to first order perturbation theory.
(Indeed there are already tantalising lattice indications,
\cite{guesken88a,sommer91a}.)

In conclusion we would just like to say that lattice results at present
are generally in qualitative agreement with other theoretical and
experimental results. However much improvement
is required to be able to make quantitative predictions.

\section{Acknowledgements}
\label{acknowledgements}

This work was supported in part by the DFG. The numerical computations
were performed on the HLRZ Cray Y-MP in J\"ulich. We wish to thank
both institutions for their support.

\end{document}